%% file: paper.tex
\documentclass[conference]{IEEEtran}

\usepackage{algorithmic, algorithm}
\usepackage[english]{babel}
\usepackage{amssymb}
\usepackage{amsmath}
\usepackage[dvipsnames]{xcolor}
\usepackage{cite}
\usepackage{subfig}
\usepackage{graphicx}
\usepackage{xspace}
\usepackage{balance}
\usepackage{amsthm}
\usepackage[hidelinks]{hyperref}
\usepackage{tikz}

\usepackage[a4paper, total={6.9in, 9.9in}]{geometry}
\normalsize

\input{macros}

\title{The Labeled Coupon Collector Problem \\ with Random Sample Sizes and Partial Recovery}
\author{\IEEEauthorblockN{Shoham Shimon Berrebi\IEEEauthorrefmark{1}, Eitan Yaakobi\IEEEauthorrefmark{1}, Zohar Yakhini\IEEEauthorrefmark{1}\IEEEauthorrefmark{2}, and Daniella Bar-Lev\IEEEauthorrefmark{3}}
\thanks{This work was funded by the European Union (DiDAX, 101115134). Views and opinions expressed are however those of the authors only and do not necessarily reflect those of the European Union or the European Research Council Executive Agency. Neither the European Union nor the granting authority can be held responsible for them.
The work of D. Bar-Lev was supported in part by Schmidt Sciences and NSF Grant CCF2212437.}
\IEEEauthorblockA{
\IEEEauthorrefmark{1}\textit{Faculty of Computer Science, Technion -- Israel Institute of Technology, Haifa, Israel}}
\IEEEauthorblockA{\IEEEauthorrefmark{2}\textit{School of Computer Science, Reichman University, Herzliya, Israel}}
\IEEEauthorblockA{\IEEEauthorrefmark{3}\textit{Center for Memory and Recording Research, University of California San Diego, CA, USA}}
\texttt{
Email: \{s-berrebi, yaakobi, zohary\}@cs.technion.ac.il,
dbarlev@ucsd.edu}
}

\IEEEoverridecommandlockouts
\IEEEaftertitletext{\vspace{-5pt}}

\begin{document}

\maketitle
\pagestyle{empty}

\begin{abstract}
We extend the Coupon Collector's Problem (CCP) and present a novel generalized model, referred as the \ekecd problem, where one is interested in recovering a bipartite graph with a perfect matching, which represents the coupons and their matching labels. We show two extra-extensions to this variation: the heterogeneous sample size case (\efmecd) and the partly recovering case.
\end{abstract}

\section{Introduction}
\label{sec:intro}
The Coupon Collector's Problem (\eccp)~\cite{intro_probability, Survey_ccp} is a probabilistic model used
for modeling in a variety of fields~\cite{ccp_survery_engineering_problems_and_computational_methods}, including statistical sampling~\cite{CCP_with_unknown_population_size_statisticsMotivated, CCP_quality_control_statistics}, algorithmic theory~\cite{Adler_Oren_Ross_2003, nonUniformCCP_application_cs_broadcast}, and network theory, where it models processes such as data collection, hashing algorithms~\cite{Adler_Oren_Ross_2003}, and even in molecular biology applications~\cite{Adler_Ahn_Karp_Ross_2005, sampling, bioCCP, CRISPR/CAS_combinatorical_design,Volf2023.11.15.567131}.
It has recently been used in the study of combinatorial coding for data storage in DNA
~\cite{goldman2013towards, anavy2019data, Sokolovskii_2024, Inbal1_ShortmersMethod, Inbal2_Coverage}.
Originally framed in terms of collecting distinct coupons from uniform and independent samples,
the \eccp studies the distribution of the number of samples required to collect all distinct coupons..
Traditionally, the \eccp involves $n$ distinct equiprobable coupons, 
and in each sample a single coupon is collected with repetitions.
In this case, the expected number of samples necessary for sampling each coupon at least once is
$n\cdot H_n\approx n\cdot \log n$, where $H_n$ is the $n$-th harmonic number.

Variants of the \eccp have emerged to model complex real-world systems.
One such variant\cite{nonUniformCCP_application_cs_broadcast} is where each coupon~$i$ has its own sampling probability $p_i$.
Another variant is the case where only $r$ distinct coupons are required~\cite{berthet2017vonschellingformulageneralized_partial_nonuniform, random_allocation_problems_FLAJOLET1992207,  leon2023ccp_partial_rth-moments, monsellato2016_partialRecovery_probabilitiesGeneratingFunctions}, instead of all the $n$ coupons.  
This problem is known as the \emph{partial \eccp} and was explored in several contexts, particularly for optimizing the collection process or estimating probabilities for subsets of the coupons.
For that variant, the expected number of samples is known to be~\cite{random_allocation_problems_FLAJOLET1992207}: \(n \cdot \sum_{i=0}^{r-1}{\frac{1}{n-i}}=n \cdot (H_n-H_ {n-r})\).
Partial recovery is also related to RAM implementations of data storage in DNA~\cite{gruica2024geometry, gruica2024combinatorial, bar2024cover}.

Another generalization of this problem is the \eccp with group drawings~\cite{betken2022_CCP_group_drawings_limits,ccp_groupdrawings_markov, stadje_group_drawings}.
This generalization considers scenarios where, instead of collecting individual coupons, in each sample,
one collects random subsets of coupons.
The size of each sample could be a constant $k$ or a random variable (RV) $K$. 
One is interested in characterizing the distribution of the number of subsets needed until each coupon was drawn in at least one of these samples.
We call these models \ccpg{K}{} or \ccpg{k}{}, depending on whether the sample size is distributed by a RV or is fixed.

Stadje~\cite{stadje_group_drawings} and Adler and Ross~\cite{Adler_Ross_2001} provide a comprehensive analysis of the \eccpg distribution,
offering bounds on the expected number of samples required to collect all coupons and efficient simulation procedures.
Stadje's work also explores applications to reliability problems, studying the case where items are sampled in fixed-size samples.
Adler and Ross studied the case
where the number of items drawn in each sample is a RV.

The \eccpg is equivalent to another problem in which the goal is to recover a graph representing matching between coupons and their labels, rather than simply collecting all coupons.
The reduction is visualized in
\autoref{fig:graph_reduction}.
Let $\Egraph = (\Coupons \,\cup\, \Labels,\, \Edges)$ be a bipartite graph with $2n$ vertices and a perfect matching. 
The vertices $C$ and $L$ represent coupons and labels.
Each sample is a set of edges between coupons and their corresponding labels. 
The goal is to recover the set $\Edges$, i.e., to identify the labels of all coupons. 
This reduced problem is equivalent to the \eccpg, as knowing the label of a coupon is equivalent to sampling it.  
It is important to keep in mind that the \eccpg is equivalent to this setup only in the case where the set of labels (i.e., the set $\Labels$) is unknown beforehand.

Following the graph setup, we define a novel sampling model, 
referred to as the \emph{Labeled Coupon Collector Problem},
in which we consider sampling sets with missing information. 
When a set of coupons is collected, its set of associated labels is also revealed, but \emph{without} the matching between coupons and labels.
The goal is to collect all coupons and recover their labels.
Note that the goal can be defined in two ways:
if the graph's vertices are known beforehand,
the last edge can be inferred,
making it equivalent to partial recovery, 
where only \( n-1 \) out of \( n \) coupons need to be collected.
Conversely,
if the vertices are not known beforehand,
the last edge cannot be inferred and must also be sampled.

In this work, we make several contributions to the study of the Labeled Coupon Collector Problem, including .
In \autoref{sec:definitions_problemStatement},
we introduce several novel extensions,
including the random sample size setting
and the partial recovery setting, which is further divided into specific recovery, targeting a predetermined subset of coupons, and non-specific recovery, where any subset of a given size suffices.
In \autoref{sec:partial_recovery},
we focus on the partial recovery extension,
studying its behavior in depth and examining the differences between specific and non-specific recovery scenarios.
In \autoref{sec:complete_recovery},
we analyze the random sample size variant,
demonstrating that the sampling process can be modeled as a Markov chain,
allowing us to calculate recovery probabilities and expected sample counts.
Across these sections, we highlight key relations between the different models.

\input{CCP_graph_reduction_figures}

\section{Definitions and Problem Statements}
\label{sec:definitions_problemStatement}
Consider a generalization of the graph presented in \autoref{fig:graph_reduction},
where instead of sampling the edges,
one only gets a set of coupons and a set of their corresponding labels,  
\emph{without knowing the matching between coupons and labels}. Formally, for an integer $n$, we let $G(n) = (C\cup L, E)$ be a bipartite graph
with $2n$ vertices and a perfect matching.
The vertices $C$ and $L$ represent the coupons and the labels.
Using this graph, we define two sampling models, namely the \ekecd and its extension the \efmecd.

\begin{definition}
\textbf{The $k$-Labeled Coupon Collector Problem (\ekecd)}. 
Given a graph $G(n)$ and an integer $k$, 
the \ekecd model considers the setup where each sample consists of a pair of subsets
$s\triangleq (s(C), s(L))$,
each of size $k$,
such that $s(C) \subset C, s(L) \subset L$,
and the sub-graph of $G(n)$ that corresponds to the nodes $s(C)\cup s(L)$ is a perfect matching.
Under this setup, the goal is to recover $E$, i.e., to match each coupon with its corresponding label.
\end{definition}

We are interested in the distribution of the number of samples needed for such a recovery.
We note that, under the graph bijection, the \ekecd differs from the \ccpg{k}{}
as instead of sampling the edges directly, we only obtain partial information about those edges.    
Similar to previous related works~\cite{ccp_groupdrawings_markov, Survey_ccp}, we show that the \ekecd can be approached using a Markov chain model.
However, as mentioned earlier,
in the \ekecd, the samples provide less information.
Specifically, each sample contains two subsets: one of $k$ coupons and another of $k$ labels, which guide the collection process but do not reveal the exact edges.

For the $\kecd{2}{}$ model we define a useful undirected graph, $H = (C, S)$, where the edges $S$ are the pairs of coupons that were sampled together, ignoring the labels.

Next, we define the Mixed Labeled Coupon Collector Problem, 
denoted as \efmecd, as an extension of the \ekecd setup,
in which a non-constant sample size is considered.
\begin{definition}
    \textbf{The Mixed Labeled Coupon Collector Problem (\efmecd)}. 
    Given a graph $G(n)$ and a RV $K$, defined by a dictionary of sample sizes and probabilities, $K \triangleq \{1:p_1,\dots,n:p_n\}$,
    wherein $P(K = k) = p_k$, $\sum_{i=1}^{n}{p_i} = 1$ and $\forall i :p_i\in[0,1]$. 
    The \efmecd model addresses a similar setup to the \ekecd setup
    but with sample sizes distributed according to $K$ instead of the constant $k$.
    \end{definition}

The \efmecd is similar to the \eccpg 
as the sample size is not necessary one.
When dealing with the special case where $P(K=1) = p$ and $P(K=2) = 1-p$, for $p\in[0,1]$, we denote the RV $K$ as $K(p) \triangleq  \{1:p,2:1-p\}$.

Given a model $M$, e.g. $M=\fmecd{K(p)}{}{n}$, we define \T{M}{r} to be the number of samples
needed to determine the labels of $r$ coupons in the model $M$. $\T{M}{n}$ is referred to as \emph{complete recovery}.
Similarly, \ST{M}{r} denotes the number of samples
needed to determine the labels of a \emph{specific} set of $r$ coupons in the model $M$ and note that from symmetry it has the value for all sets of the same size.
Note that $\T{M}{n}=\ST{M}{n}$, and therefore we only 
consider the differences when discussing partial recovery, that is, the case where $r < n$.
\Min{M}{r} is defined as the minimal value that the RV \T{M}{r} can attain.
For all notations above we assume a total of $n$ coupons.
Moreover, if the model can be inferred from the context,
or if the relevant statement holds for every model,
we may omit $M$, i.e., \T{}{r}.

\begin{example}
\label{example_defintions}
    Consider $M = \fmecd{K(p)}{}{5}$ with $n=5$ coupons \( C = \{A, B, C, D, E\} \) and their labels' matching $\{A:1,B:2,C:3,D:4,E:5\}$.
    We sample with probability $p$ each of the five coupons and its label, 
    and in the complementary event we sample $2$ coupons,
    w.l.o.g. $\{A,B\}$ and their labels $\{1,2\}$,
    without knowing whether the matching is $\{A:1,B:2\}$ or $\{A:2,B:1\}$.
    $\T{M}{5}$ will be the number of samples until recovery of the full matching, 
    meanwhile, $\T{M}{1}$ is  for recovery of any one coupon and its matching label
    and $\ST{M}{1}$ is for recovery of a specific (w.l.o.g. $\{A:1$\}) coupon and its label.
\end{example}

We now present our problem statements.

Inspired by existing extensions of the \eccp problem, we are also
interested in scenarios where we wish to know the labels of only a subset of the coupons.
This leads us to define the following, representing partial recovery questions:

\begin{problem}[Partial Coverage in the \ekecd]
\label{Problem:Partial Coverage}
Consider $M = \ekecd$ wherein $k\in[n]$.
We wish to solve the following problems for any integer $0 < r < n$:
\begin{enumerate}
    \item \textbf{Arbitrary subset:} 
    Find $\Min{M}{r}$
    and $\E[\T{M}{r}]$.
    \item \textbf{Specific subset:} 
    Find $\E[\ST{M}{r}]$.
\end{enumerate}
\end{problem}

In the \ccpg{k}{}, determining all coupon labels requires at least $\lceil {\frac{n}{k}} \rceil$ samples, however, in the \ekecd, when acquiring a coupon does not guarantee knowing its label, $n$ samples may not suffice. Moreover, usually the best-case scenario is not occurring; therefore, we are interested in the distribution.

\begin{problem}[Complete Recovery in the \efmecd]
\label{Problem:complete_recovery}
    Let $K$ be a RV and $M = \fmecd{K}{}{}$,
    we wish to solve the following:
    \begin{enumerate}
    \item \textbf{Minimal number of samples}\label{Problem:Minimal number:(k1,k2)-MECD} 
    Find $\Min{M}{n}$. 
    \item \textbf{$\T{M}{n}$ Distribution:} \label{Problem:Expected number}
    Find $\E[\T{M}{n}]$ and for any $t \in \N$, find $P(\T{M}{n}>t)$.
    \end{enumerate}
\end{problem}

This paper mainly addresses the \efmecd variant in the context of the ``complete recovery'' problem and the \ekecd variant in the context of the ``partial recovery'' problem.

\input{partial_recovery_section}
\input{complete_recovery_section}

\section{Discussion}
\note{do we want to say anything on "no convex solution" :
The expected number of samples for recovery in the \fmecd{K}{}{} 
is not simply a convex combination of the \kecd{k}{} models from them it complex.
Therefore}
The representation of \fmecd{K}{}{} as a Markov chain has strong impact on the use of this model,
as it shows equivalence between the \efmecd model and a known model with well-known algorithms.
It paves the way for exact calculation
of expected number of samples for recovery and other values of interest
in any \efmecd model 
once we extend the Markov approach for \kecd{k}{} for $k > 2 $.

Future directions include examining the conjectures stated in this paper, as well as further extending the model.
In particular, it may be interesting to consider a set of labels for each coupon
and to investigate the sampling complexity of collecting subsets of these labels.

\section{Acknowledgments}
We thank the Yakhini and Yaakobi research groups for important and useful discussions. We specifically thank Oriel Limor for valuable guidance and explanations related to the Markov chain representation.
We also thank Yoav Danieli, Elad Kinsbruner and Lior Shiboli for useful discussions and comments.

\newpage
\balance
\bibliographystyle{IEEEtran}

\clearpage

\end{document}

%% file: macros.tex
\theoremstyle{definition}
\newtheorem{theorem}{Theorem}
\newtheorem{lemma}{Lemma}

\newtheorem{definition}{Definition}
\newtheorem{conjecture}{Conjecture}
\newtheorem{example}{Example}

\newtheorem{problem}{Problem}
\newtheorem{claim}{Claim}

\addto\extrasenglish{

}

\newcommand{\N}{\mathbb{N}}

\newcommand{\E}{\mathbb{E}}

\def\CC#1#2{\ensuremath{\left(\kern-.3em\left(\genfrac{}{}{0pt}{}{#1}{#2}\right)\kern-.3em\right)}}

\newcommand{\note}[1]{}

\newcommand{\eccp}{\ensuremath{\text{CCP}}\xspace}
\newcommand{\eccpg}{\ensuremath{K\text{-CCP}}\xspace}
\newcommand{\ccp}[1]{\ensuremath{\text{CCP}}\xspace}
\newcommand{\ccpg}[2]{\ensuremath{#1\text{-CCP}}\xspace}

\newcommand{\ekecd}[1][k]{\ensuremath{#1\text{-LCCP}}\xspace}
\newcommand{\kecd}[2]{\ensuremath{#1\text{-LCCP}}\xspace}

\newcommand{\fmecd}[3]{\ensuremath{{#1}\text{-LCCP}}\xspace}
\newcommand{\efmecd}{\ensuremath{K\text{-LCCP}}\xspace}

\newcommand{\T}[2]{\ensuremath{T_{#1}(#2)}\xspace} 
\newcommand{\ST}[2]{\ensuremath{S_{#1}(#2)}\xspace} 
\newcommand{\Min}[2]{\xspace\ensuremath{\mathrm{Min}_{#1}(#2)}\xspace} 

\newcommand{\ns}{s'}

\newcommand{\Egraph}{\ensuremath{G}\xspace}
\newcommand{\Coupons}{\ensuremath{C}\xspace}
\newcommand{\Labels}{\ensuremath{L}\xspace}
\newcommand{\Edges}{\ensuremath{E}\xspace}

\let\oldtextbf\textbf
\renewcommand{\textbf}[1]{{\boldmath\oldtextbf{#1}}}

%% file: CCP_graph_reduction_figures.tex
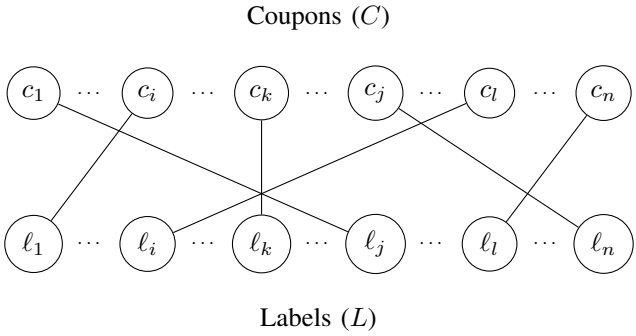
\begin{figure}
\centering

    \centering
    \begin{tikzpicture}[scale=0.5, auto, swap]
    
        \node[draw, circle] (c1) at (-9, 2) {$c_1$};
        \node[scale=0.75] at (-7.5, 2) {$\cdots$};
        \node[draw, circle] (ci) at (-6, 2) {$c_i$};
        \node[scale=0.75] at (-4.5, 2) {$\cdots$};
        \node[draw, circle] (ck) at (-3, 2) {$c_k$};
        \node[scale=0.75] at (-1.5, 2) {$\cdots$};
        \node[draw, circle] (cj) at (0, 2) {$c_j$};
        \node[scale=0.75] at (1.5, 2) {$\cdots$};
        \node[draw, circle] (cl) at (3, 2) {$c_l$};
        \node[scale=0.75] at (4.5, 2) {$\cdots$};
        \node[draw, circle] (cn) at (6, 2) {$c_n$};
    
        \node[draw, circle] (v1) at (-9, -2) {$\ell _1$};
        \node[scale=0.75] at (-7.5, -2) {$\cdots$};
        \node[draw, circle] (vi) at (-6, -2) {$\ell _i$};
        \node[scale=0.75] at (-4.5, -2) {$\cdots$};
        \node[draw, circle] (vk) at (-3, -2) {$\ell _k$};
        \node[scale=0.75] at (-1.5, -2) {$\cdots$};
        \node[draw, circle] (vj) at (0, -2) {$\ell _j$};
        \node[scale=0.75] at (1.5, -2) {$\cdots$};
        \node[draw, circle] (vl) at (3, -2) {$\ell _l$};
        \node[scale=0.75] at (4.5, -2) {$\cdots$};
        \node[draw, circle] (vn) at (6, -2) {$\ell _n$};
    
        \draw[-] (c1) -- (vj);
        \draw[-] (cn) -- (vl);
        \draw[-] (ci) -- (v1);
        \draw[-] (cj) -- (vn);
        \draw[-] (cl) -- (vi);
        \draw[-] (ck) -- (vk);
    
        \node at (-1.5, 4) {Coupons ($C$)};
        \node at (-1.5, -4) {Labels ($L$)};
        
    \end{tikzpicture}
    

\caption{\footnotesize
CCP by collecting edges in a bipartite graph. The coupons (\(C\)) are matched to the labels (\(L\)) via edges, representing samples. This graph-based representation highlights the CCP as a problem of reconstructing all edges with minimal sampling.\label{fig:graph_reduction}
}
\vspace{-4ex}
\end{figure}

%% file: partial_recovery_section.tex
\section{\texorpdfstring{\hyperref[Problem:Partial Coverage]{Partial Recovery in the \ekecd[k]}}{Partial Recovery in the \ekecd[k]}}
\label{sec:partial_recovery}
We start by noting that the \kecd{1}{n} is equivalent to the standard \ccp{n} and therefore 
$\T{\kecd{1}{n}}{r}$ and $\ST{\kecd{1}{n}}{r}$ have the same distributions as their known standard counterparts.

We now address a connection between different \ekecd models.
\begin{lemma}
\label{lemma:n-k and k}
    For any integers $ 0 < k < n$ it holds that $\T{\kecd{k}{n}}{n}$ and $\T{\kecd{(n-k)}{n}}{n}$ have the same distribution.
\end{lemma}
\begin{proof}
    Denoting the problem's graph as $G=(C,V)$. We can use the same graph for both models as the amount of coupons is equal in both models.
    Let  $S=(S_C,S_V)=(\{c_1,\dots,c_k\},\{v_1,\dots,v_k\})$ be a sample in the \kecd{k}{n} model,
    such that $C$ is the sampled coupons and $V$ are their labels.
    We can define the equivalent sample in the \kecd{(n-k)}{n} model as $S'=(S'_C,S'_V)=(\{c_{k+1},\dots,c_n\},\{v_{k+1},\dots,v_n\})$.
    The probabilities for both samples is equal as $\binom{n}{n-k}=\binom{n}{k}$ and they give the same information if we know the graph vertices beforehand, as one can easily calculate $(S'_C,S'_V) = (S'\setminus C,V \setminus S'_V)$.
    For complete recovery, we need to sample each vertex at least once, so we surely have $(C,V)$.
\end{proof}

We turn to the \kecd{2}{} model, starting with a useful claim.
\begin{lemma}
\label{claim:knowing_c__size_3}
    Let $G(n) = (C\cup L, E)$ and $H=(C,S)$ graphs of \kecd{2}{}.
    We know the label of coupon $c\in C$, $L(c)$, if and only if $c$ is part of a connected component of size at least $3$ in the graph $H$. 
\end{lemma}

\begin{proof}   
We prove this considering both directions:

\begin{itemize}
    \item[$\Rightarrow$] 
    Suppose, for contradiction, that $c$ belongs to a component of size less than 3. There are two possibilities:
    \begin{enumerate}
        \item $c$ is an isolated coupon. In this case, $c$ has not been sampled, and thus we cannot deduce its label.
        \item $c$ is connected to exactly one other coupon. This is also insufficient to determine its label since the sample gives the 2 labels and 2 coupons without matching them.
    \end{enumerate}
    In both cases, we cannot deduce $L(c)$, contradicting the assumption that $L(c)$ is known. Therefore, if we know $L(c)$, $c$ must be part of a component of size at least 3.
    
    \item[$\Leftarrow$] 
    Now assume $c$ belongs to a connected component of size at least 3. We can deduce the label of $c$ by considering the following:
    
    Let $c'$ be a vertex in the same component as $c$ such that $d(c') \geq 2$. This means $c'$ appears in at least two samples, each containing different coupons. From this information, we can deduce the label of $c'$.
    
    Since the component is connected, there exists a path from $c'$ to $c$. After deducing the label of $c'$, we can propagate this information along the path to $c$, allowing us to infer $L(c)$ as well.
\end{itemize}

Thus, the claim is proved.
\end{proof}

As a result of \autoref{claim:knowing_c__size_3} we deduce the two following claims, 
in the context of any total number of coupons, $n$.
\begin{claim}
In the \kecd{2}{} model, recovering the first coupon implies the recovery of two additional coupons. That is,
    \vspace{-1ex}
    $$\T{\kecd{2}{n}}{r=1}  = \T{\kecd{2}{n}}{r=2} = \T{\kecd{2}{n}}{r=3}.$$
\end{claim}

\begin{claim}
\label{claim:minimum_2ECD}
In the \kecd{2}{} model, one must sample at least $\left\lceil \frac{2n}{3}\right\rceil$ for complete recovery:
$\Min{\kecd{2}{n}}{n} = \left\lceil \frac{2n}{3}\right\rceil$.
\end{claim}
This follows by observing that the number of edges (samples) in a graph $H$ 
is at least the total number of vertices (coupons) minus the number of connected components. 
For $H$ with $\left\lfloor \frac{n}{3}\right\rfloor$ components partition the vertices into triplets \note{last - 4 or 5} and put two edges in each.
\note{full proof in proposal}

\begin{theorem}
\label{theorem:expected_2ecd_specific_partialrecover_1}

For recovering a specific coupon in the \kecd{2}{} model, it holds that
$$\E[\ST{\kecd{2}{n}}{1}]  = \frac{n \left(n^3 + n^2 + 5n - 5\right)}{\left(n + 3\right)\left(5n - 4\right)}
= O(n^2).$$
\end{theorem}
\begin{proof}
    Let $t \in \mathbb{N}$ and denote by $c$ the specific coupon we aim to recover.
    We are interested in
    $P(\ST{\kecd{2}{n}}{1}>t)$
    because 
    $\E[\ST{\kecd{2}{n}}{1}] = \sum_{t=0}^{\infty}P(\ST{\kecd{2}{n}}{1}>t)$.
    One can notice that there are $2$ distinct events in which $c$ isn't recovered after $t$ samples:
    \begin{itemize}
        \item Event A: 
        We did not see the coupon $c$ in the $t$ draws.
        The probability is
        $P(A)=\frac{\binom{n-1}{2} ^t }{\binom{n}{2} ^t} = \left(\frac{n-2}{n}\right)^t$.
        \item Event B: 
        We have seen the coupon $c$ but not able to recover its label.
        This scenario happens if and only if the coupon only appeared with another coupon $c'$
        which itself did not appear with any other coupon.
        Denote by $\ell$ the number of times the sample $\{c,c'\}$ appeared.
        There are $n-1$ options to choose the coupon $c'$,
        $\binom{n-2}{2}$ options to choose the coupons for each of the $t-\ell$ samples 
        and $\binom{t}{\ell}$ options to choose which of the samples is $\{c,c'\}$.
        We can sum over all the options for $\ell$ and use the binomial expansion to get:
        $
        P(B) 
        =
        \frac{\sum_{\ell=1}^t {\binom{n-2}{2}^{t-\ell} \cdot \binom{t}{\ell} \cdot (n-1)}}{\binom{n}{2} ^t} 
        = 
        \frac{\left(\binom{n-2}{2} + 1 \right)^t - \binom{n-2}{2}^t}{\binom{n}{2} ^t} \cdot (n-1) 
        = 
        \left(\frac{(n-2)(n-3)+2}{n(n-1)}\right) ^t \cdot (n-1)
        -
        \left(\frac{(n-2)(n-3)}{n(n-1)}\right) ^t \cdot (n-1)
        $.
    \end{itemize}
    We denote $q_1 = \frac{n-2}{n}$,  $q_2 = \frac{(n-2)(n-3)+2}{n(n-1)}$ and $q_3 = \frac{(n-2)(n-3)}{n(n-1)}$.
    Since $A$ and $B$ are disjoint:
    \[\begin{array}{lcl}
    \hspace{-0.06cm}
    P(\ST{\kecd{2}{n}}{1}>t)  
    \hspace{-0.08cm}=\hspace{-0.08cm}
    P(A) \hspace{-0.05cm}+\hspace{-0.05cm} P(B) 
    \hspace{-0.08cm}=\hspace{-0.08cm}
    q_1^t
    \hspace{-0.02cm}+\hspace{-0.02cm}
    (q_2 ^t
    \hspace{-0.05cm}-\hspace{-0.05cm}
    q_3 ^t)
    (n\hspace{-0.05cm}-\hspace{-0.05cm}1)
    \end{array}
    \]
    Therefore,
    \begin{align*}
    \E[\ST{\kecd{2}{n}}{1}] 
    &=
    \sum_{t=0}^{\infty}
    \left(
    q_1^t
    +
    q_2 ^t (n-1)
    -
    q_3 ^t (n-1)
    \right)
    \\&=
    \frac{n}{2}
    +
    \frac{n(n-1)^2}{5n-4}
    -
    \frac{n(n-1)}{2(n+3)}
    \\&=
    \frac{n \left(n^3 + n^2 + 5n - 5\right)}{\left(n + 3\right)\left(5n - 4\right)} 
     \vspace{-2ex}
    \end{align*}
     \vspace{-1.5ex}
\end{proof}
\begin{theorem}
\label{theorem:expected_2ecd_nonSpecific_partialrecovered_1}
Denoting $m = (n-2)(n+1)/2$, for recovering any coupon in the \kecd{2}{}:
$$\E[\T{\kecd{2}{n}}{1}]  
        = 
        1
        +
        \sum_{i=1}^{\frac{n}{2}}
            \frac{n!}{(n-2i)! 2^i}
            \frac{\left(m- i\right)!}{m!}
        = 
        O(n)
            $$
\end{theorem}
The proof is similar to the proof of \autoref{theorem:expected_2ecd_specific_partialrecover_1}
and given in the journal version.
\note{the proof is commented}

%% file: complete_recovery_section.tex
\section{\texorpdfstring{\hyperref[Problem:complete_recovery]{Complete Recovery in the \efmecd}}{Complete Recovery in the \efmecd}}
\label{sec:complete_recovery}

Now, we shift our focus to the expected number of samples needed for complete recovery.

\begin{theorem}
    Let a RV $K$ be the sample size distribution 
    in both the \fmecd{K}{}{} and the \ccpg{K}{},
    the following holds:
    \begin{equation}
    \label{theorem:expected:MECD_vs_CCP}
        \E[\T{\fmecd{K}{P}{n}}{r=n}] > \E[\T{\ccpg{K}{n}}{r=n-1}].
    \end{equation}
\end{theorem}

\begin{proof}
Let \fmecd{K}{P}{n}
and $\ccpg{K}{n}$ be two models with the same sample size distribution, $K$.
Also, $\mathcal{P}(C)$ be the set of all possible samples of $C$.
Moreover, $\mathcal{S}(\mathcal{P}(C))$ is the set of all possible finite sequences of samples.
Every sequence with $i$ samples $A_i \in \mathcal{S}(\mathcal{P}(C))$ may recover one of the models, both or none.
To show \autoref{theorem:expected:MECD_vs_CCP}, we are interested in showing the following:

\[\forall i  \in \mathbb{N} ; P(\T{\fmecd{K}{P}{n}}{n} \leq i) \leq P(\T{\ccpg{K}{n}}{n-1} \leq i)\]
\[\exists j \in \mathbb{N} ; P(\T{\fmecd{K}{P}{n}}{n} \leq j) < P(\T{\ccpg{K}{n}}{n-1} \leq j)\]

Let $M \in \{\efmecd, \eccpg\}$. 
We say that $A\in M(r)$ 
if and only if 
the event $A$ in the model $M$ suffices a recovery of $r$ coupons. 
Now, we can separate to disjoint events: 
\[ P(\T{\fmecd{K}{P}{n}}{n} \leq i) =
\hspace{-0.6cm}
\sum_{A_i \in \fmecd{K}{P}{n}(n)} \hspace{-0.6cm}  P(A_i)
\]
\vspace{-1ex}
\[ P(\T{\ccpg{K}{n}}{n-1} \leq i) =
\hspace{-0.6cm} 
\sum_{A_i \in \ccpg{K}{n}(n-1)} \hspace{-0.6cm} P(A_i)
\]

The probabilities of the events are the same in both models as they share the same sample size distribution.
Therefore, we can use the graph equivalence of \autoref{fig:graph_reduction} 
between \ccpg{K}{n} and \fmecd{K}{P}{n} 
and notice that if an event $A_i$ recovers all the $n$ edges in \fmecd{K}{P}{n} 
it means we collected at least $n-1$ edges (and may have inferred the last one) 
and therefore recovered at least $n-1$ coupons in the \ccpg{K}{n} model.
Therefore:
\vspace{-1ex}
\begin{align*}
P(\T{\fmecd{K}{P}{n}}{n} \leq i) = & 
\sum_{A_i \in \fmecd{K}{P}{n}(n)} P(A_i) \\
\leq & \sum_{A_i \in \ccpg{K}{n}(n-1)} P(A_i)\\
= & P(\T{\ccpg{K}{n}}{n-1} \leq i)
\end{align*}

Now, we show that the inequality is strict by looking at the minimal number of samples for recovery in both models.
Denote $k_m=\arg\min_{k}\{|k - \frac{n}{2}| : P(K=k)>0\}$.
We now deduce $\Min{\ccpg{K}{n}}{n-1} \leq \left\lceil\frac{n-1}{k_m}\right\rceil$.

Since $\Min{\fmecd{K}{P}{n}}{n} = \Min{\kecd{k_m}{n}}{n} = \left\lceil\frac{2}{k_m+1} \cdot n \right\rceil >
\left\lceil\frac{n-1}{k_m}\right\rceil \geq \Min{\ccpg{K}{n}}{n-1}$,
we can infer $\Min{\fmecd{K}{P}{n}}{n} > \Min{\ccpg{K}{n}}{n-1}$, 
meaning that there is an event $A$ that recovers $n-1$ coupons in \ccpg{K}{n} 
but does not recover $n$ edges in \fmecd{K}{P}{n}.

This implies $P(\T{\fmecd{K}{P}{n}}{n}\hspace{-0.018cm} \leq \hspace{-0.018cm} i) < P(\T{\ccpg{K}{n}}{n-1} \hspace{-0.018cm} \leq \hspace{-0.018cm} i)$, and
therefore,
        $\E[\T{\fmecd{K}{P}{n}}{n}] >  \E[\T{\ccpg{K}{n}}{n-1}]$. 
\end{proof}


\label{Markov:Finite-MECD}
We now shift our focus to the practical aspects of working with these models, demonstrating their representation as a Markov process.
We have already seen the sampling process in the \eccpg~\cite{Survey_ccp} and the \kecd{2}{n}~\cite{barlev2024} models
can be approached using Markov chains,
implying a process for calculating $\E[\T{\kecd{2}{n}}{n}]$.
We extend this to the \fmecd{K(p)}{}{n} model to account for the variability in random sample sizes,
extending the approach to calculating $\E[\T{\fmecd{K(p)}{}{n}}{n}]$.
\begin{theorem}
For any $p \in [0,1]$ the $\fmecd{K(p)}{}{n}$ model can be represented as a Markov chain. 
\label{Markov:(1;2)-MECD}
\end{theorem}
\begin{proof}
In the $\fmecd{K(p)}{}{n}$ model, each state of the Markov chain represents a configuration of collected coupons.  
We define the state as $s = (\alpha, \beta, \gamma)$, where:
$\alpha$ is the number of ``unknown'' coupons,
$\beta$ of ``partly known'' coupons (collected but labels still unknown) and
$\gamma$ of ``known'' coupons (collected and identified).
To summarize, the Markov chain state space is
$\Sigma = \{(\alpha, \beta, \gamma) \mid \alpha + \beta + \gamma = n, \, \alpha, \beta, \gamma \in [0,n],  \beta \text{ is even}\}$,
and the transition matrix $M$, where $M_{s \to \ns}$ gives the probability of transitioning from state $s = (\alpha, \beta, \gamma)$ to $\ns = (\alpha', \beta', \gamma')$.

The transition matrix is calculated by considering all possible samples under the abstraction of the Markov process.
Using \autoref{claim:knowing_c__size_3}, $(\alpha, \beta, \gamma)$ are numbers of vertices in connected components of sizes $(1,2, \geq3)$.
Sample of size $1$ is viewed as moving this coupon to a connected component of size $3$, 
even if there is no such connected component.
One can add $3$ coupons to the system, assume they are already known, connected to each other and connect a coupon sampled alone to it.
We denote
$\mu = \frac{1-p}{\binom{n}{2}}, \lambda = \frac{p}{n}$,
as the coefficients for probability normalizing when counting the number of possible samples.

\vspace{-2.5ex}
\begin{equation}
\label{transistion_matrix_markov_(1;2)-MECD}
M_{s \to \ns}
\hspace{-0.1cm}=\hspace{-0.1cm}
\begin{cases}
\mu \binom{\alpha}{2}, & 
\ns = (\alpha - 2, \beta + 2, \gamma)\\
\mu \alpha \beta, &
\ns = (\alpha - 1, \beta - 2, \gamma + 3) \\
\mu \cdot 4 \binom{\beta/2}{2}, &
\ns = (\alpha,\beta - 4, \gamma + 4) \\
\mu \alpha \gamma + \lambda \alpha, & 
\ns = (\alpha - 1,\beta, \gamma + 1) \\
\mu \beta \gamma + \lambda \beta, & 
\ns = (\alpha, \beta - 2, \gamma + 2) \\
\mu \left(\binom{\gamma}{2} + \frac{\beta}{2}\right) + \lambda \gamma, & 
\ns = (\alpha, \beta, \gamma) \\
\ 0, & \text{otherwise}
\end{cases}
\end{equation}
To compute the expected number of samples required to collect all edges,
we analyze the hitting times of this Markov chain.
Specifically, we calculate the expected number of transitions (representing samples)
needed to reach the absorbing state $j=(0,0,n)$,
where all edges are known.
Standard Markov chain theory techniques are used to compute hitting times,
applying the recursive equation:
\[
E_{ij} = \begin{cases} 
0, & \text{if } i = j \\
\frac{1}{1 - P_{ii}} \left( 1 + \sum_{k \neq i} P_{ik} E_{kj} \right), & \text{if } i \neq j 
\end{cases}
\]
\end{proof}
\begin{claim}
    Let a RV $K$ be the sample size distribution. 
    The $\fmecd{K}{}{n}$ model can be represented as a Markov chain.
\end{claim}
As above, we do this by adding complexity to the fixed \ekecd[k] process.
In each transition the sample size is determined as a first step and then the transition probabilities of the relevant \ekecd[k] can be used, as a second step.
Namely:
\begin{enumerate}
    \item 
    Sample Size Decision: From the current 
    ``old state'', the process transitions to one of $\frac{n}{2}$ potential states, each representing a specific sample size.
    \item 
    Coupon Collection Transition: After the sample size is determined, the system evolves as in the \ekecd[k] Markov process, updating the different ``partly known'' counters (like $\beta$ in the proof for $\fmecd{K(p)}{}{}$, but more).
\end{enumerate}

\begin{figure}[t]
    \centering
    \vspace{-2.6ex}
    \includegraphics[width=1\linewidth]{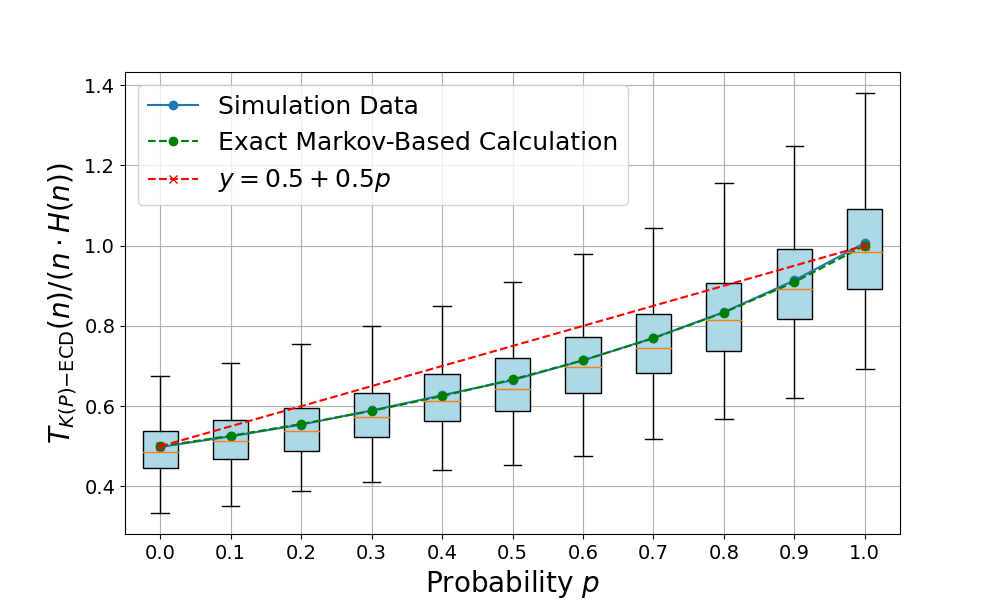}
    \caption{\footnotesize The normalized value of $\T{\fmecd{K(p)}{}{n}}{n}$ as a function of $p\in[0,1]$, for $n=2000$. 
    For any $p \in \{0,0.1,\ldots,1\}$, the exact value of $\E[\T{\fmecd{K(p)}{}{n}}{n}]$, 
    calculated by the Markov approach (\autoref{transistion_matrix_markov_(1;2)-MECD})
    is presented by the green curve, 
    while the dashed red line represents the convex combination.
    Additionally, we conducted simulations of $\T{\fmecd{K(p)}{}{n}}{n}$ and present, as box plots, results for $1000$ instances per all relevant values of $p$.\vspace{-4.5ex}}
    \label{fig:avgTbyP}
    \end{figure}

Using the Markov approach,
we demonstrate that there is no simple convex solution to $\E[\T{\fmecd{K(p)}{}{n}}{n}]$.
According to our simulations and based on exact calculations for any specific $n$,
we conjectured that $\E[\T{\kecd{2}{n}}{n}] \approx \frac{n H_n}{2}$.
This is also independently observed in a parallel work~\cite{barlev2024}.
\label{NoSimpleConvexSolution}
Building on this conjecture, intuition may suggest that
        \[\begin{array}{lcl}
        \hspace{-0.16cm}
        \E[\T{\fmecd{K(p)}{}{n}}{n}] 
        &\hspace{-0.28cm}\stackrel{?}{=}&
        \hspace{-0.28cm}p \E[\T{\kecd{1}{n}}{n}] \hspace{-0.08cm} + \hspace{-0.08cm} (1-p) \E[\T{\kecd{2}{n}}{n}] 
        \\&\hspace{-0.28cm}\stackrel{?}{\approx}&
        \hspace{-0.28cm}p nH_n + (1-p) \frac{n H_n}{2}
        =
        p \frac{nH_n}{2}+\frac{nH_n}{2}
        \end{array}\]
    For a fixed number of coupons, the expression $nH_n$ is a constant, and therefore we would expect a linear relationship between $\E[\T{\fmecd{K(p)}{}{n}}{n}]$ and $p$. 
    However, as seen in \autoref{fig:avgTbyP} this is not typically the situation.
\\
\autoref{theorem:K(p)-ECD(3)} provides a closed form solution for $n=3$.
\begin{theorem}
    \label{theorem:K(p)-ECD(3)}
    For any $p\in[0,1]$,
    we have that
    $\E[\T{\fmecd{K(p)}{}{3}}{r=n=3}] =
    \frac{13p^4+63p^3+74p^2-200p-120}{4(p+2)^2(3-p)}$.
\end{theorem}
\begin{proof}
Let $T \triangleq \T{\fmecd{K(p)}{}{3}}{3}$ for simplicity. 
We have that:
\[
\E[T] \hspace{-0.04cm}= \hspace{-0.08cm}\sum_{t=0}^\infty P(T > t) \hspace{-0.04cm}= \hspace{-0.08cm}\sum_{t=0}^\infty \sum_{s = 0}^t P(S=s) P(T > t \mid S\hspace{-0.04cm}=\hspace{-0.04cm}s)
\]
where $0 \leq S \leq T$ governs the number of samples of size-$1$ obtained in the first $t$ samples.

Note that $P(s) = \binom{t}{s} p^s (1-p)^{t-s}$ 
and the boundary cases:
\begin{itemize}
    \item $P(T > 0) = P(T > 1) = 1$.
    \item $P(T > t \mid S = 0) = \frac{1}{3^{t-1}}, P(T > t \mid S = 1) = \frac{3}{3^{t-1}}.$
    \item $P(T > t \mid S = t) = \frac{t}{3^{t-1}}.$
\end{itemize}

Additionally, for $t > 1$ and $1 < S < t$ there exist two cases where we fail to recover the graph with $3$ coupons.

\textbf{Case A:} All the $S$ samples of size-$1$ are of one specific coupon 
and all the $t-S$ samples of size-$2$ are of the same pair of $2$ coupons 
(this pair of coupons may contain the unique coupon that corresponds with the samples of size-$1$).
Thus, the relevant probability is $P(A^{t,s}) 
\triangleq
P(A \mid T=t, S=s)
= 
\frac{3^2}{3^t}.$

\textbf{Case B:} All $S$ samples of size-$1$ are of two specific coupons (excluding a specific coupon).
All $t-S$ samples of size-$2$
are of the same pair that contains the $2$ coupons allowed in the samples of size-$1$.
Note that there is freedom regarding how the $S$ samples are divided between the coupons. 
Thus, the relevant probability is
$
P(B^{t,s}) 
\triangleq 
P(B \mid T=t, S=s) 
=
\frac{3s}{3^t}.
$

Combining the cases for $t > 1$ and $1 < S < t$:
\vspace{-1ex}
\[
P(T > t \mid S = s) 
=
P(A^{t,s}) 
+
P(B^{t,s}) 
=
\frac{s +3}{3^{t-1}}.
\]

Finally, summing over all cases, we obtain:
\[\begin{array}{l@{\hspace{0.4em}}c@{\hspace{0.4em}}l}
\mathbb{E}[T] 
&=& 
\sum_{t=0}^\infty P(T > t) 
\\& = &
2
+ 
\sum_{t=2}^\infty 
    \left(
    \sum_{s=2}^{t-1}
    \binom{t}{s} p^s (1-p)^{t-s} \frac{s +3}{3^{t-1}}
        \right. \\&& \left. \hspace{1.8cm}
        + \frac{(1-p)^t}{3^{t-1}} + \frac{t p (1-p)^{t-1} \cdot 3}{3^{t-1}} + \frac{p^t t}{3^{t-1}} 
    \right)
\\&\stackrel{(*)}{=}& 
\frac
{13p^4 + 63p^3 + 74p^2 - 200p - 120}
{4(p+2)^2(3-p)},
\end{array} 
\]
where (*) in the last equality follows from basic algebra and known series sums.
\end{proof}

Already knowing that $\Min{\kecd{2}{n}}{n} = \left\lceil \frac{2n}{3}\right\rceil$ (\autoref{claim:minimum_2ECD}) and $\Min{\kecd{1}{n}}{n} = n$ (sampling each coupon once), a question arises as to whether allowing samples of both models, from both sizes, reduces the minimal number of samples needed for a complete recovery. We show that for $n \geq 3$ this value in \fmecd{K(p)}{}{n} inherits the value in the \kecd{2}{n} model, formally, for any $n \geq 3$, we have that: 
$$
\Min{\fmecd{K(p)}{}{n}}{n} = \Min{\kecd{2}{n}}{n} = \left\lceil \frac{2n}{3}\right\rceil.
$$
Moreover, we conjectured that the minimal number of samples needed for complete recovery in \ekecd follows a linear relationship with $\frac{1}{k+1}$:
$
        \Min{\kecd{k}{n}}{n} = \left\lceil\frac{2}{k+1} \cdot n \right\rceil 
$.
\note{the proof for \kecd{2}{n} is commented}

This conjecture
was proven for values of $n$ that are divisible by $\binom{k+1}{2}$ in a parallel work~\cite{barlev2024}.
However, proving or refuting it for any $n > k$,  remains an open problem.

We extended this result and present the behavior in the \efmecd model and its relation to the \ekecd.
\begin{theorem}
\label{conjecture:minimal:MECD}
    Let a RV $K$ be the sample size distrubtion.
    The following holds:
    $\Min{\fmecd{K}{}{n}}{n} = \Min{\kecd{k_m}{n}}{n}$,
    where $k_m=\arg\min_{k}\{|k - \frac{n}{2}| : P(K=k)>0\}$.
\end{theorem}
Note that
$\arg\min_{k}\{|k - \frac{n}{2}| : P(K=k)>0 \}$
is equal to
$\arg\min_{k}\{\Min{\kecd{k}{n}}{n} : P(K=k)>0\}$
using the conjecture on the minimal number of samples needed for complete recovery in \ekecd and the \autoref{lemma:n-k and k}.
This theorem is proved using the Pigeonhole Principle and the proof is available in the journal version.

From the relation between the \ekecd and \ccpg{k}{} and the behavior of \Min{\kecd{k}{}}{n}, 
we conjecture a relationship between \ekecd[k] with different sample sizes and the general case \efmecd.

\begin{conjecture}
\label{conjecture:expected:(k1,k2)-MECD relation with k-ECD}
    Let $K = \{k_1:p, k_2:1-p\}$ and assume w.l.o.g. $k_1 < k_2 \leq \frac{n}{2}$. 
    The models $\fmecd{K}{}{n}$, \kecd{k_1}{n} and \kecd{k_2}{n} satisfy:
    $$
    \E[\T{\kecd{k_2}{n}}{n}] \leq \E(\T{\fmecd{K}{}{n}}{n}) \leq \E[\T{\kecd{k_1}{n}}{n}].
    $$
\end{conjecture}
Note that if $\max(k_1,k_2) > \frac{n}{2} $, \autoref{lemma:n-k and k} can be used.